# Half-metallicity in honeycomb-kagome-lattice Mg$_3$C$_2$ monolayer with carrier doping

Hongzhe Pan[1,2*], Yin Han[1], Jianfu Li[2], Hongyu Zhang[3], Youwei Du[1] and Nujiang Tang[1*]

[1]*National Laboratory of Solid State Microstructures, Collaborative Innovation Center of Advanced Microstructures and Jiangsu Provincial Key Laboratory for Nanotechnology, Nanjing University, Nanjing 210093, China*

[2]*School of Physics and Electronic Engineering, Linyi University, Linyi 276005, China*

[3]*Department of Physics, East China University of Science and Technology, Shanghai 200237, China*

E-mail: panhongzhe@lyu.edu.cn, tangnujiang@nju.edu.cn

## Abstract

To obtain high-performance spintronic devices with high integration density, two-dimensional (2D) half-metallic materials are eagerly pursued all along. Here, we propose a stable 2D material with a honeycomb-kagome lattice, i.e., the Mg$_3$C$_2$ monolayer, based on the particle-swarm optimization algorithm and first-principles calculations. This monolayer is an anti-ferromagnetic (AFM) semiconductor at its ground state. And we demonstrate that a transition from AFM semiconductor to ferromagnetic half-metal in this 2D material can be induced by carrier (electron or hole) doping. The half-metallicity arises from the 2p$_z$ orbitals of the carbon (C) atoms for the electron-doped system, and from the C $2p_x$ and $2p_y$ orbitals for the case of hole doping. Our findings highlight a new promising material with controllable magnetic and electronic properties toward 2D spintronic applications.

## 1. Introduction

As a high-speed and low-energy-consuming information technology, spintronics is considered to be the most promising candidate to replace conventional electronics, and thus has attracted tremendous attentions from both science and industry during the past few decades [1]. With the development of spintronics, developing new spintronic materials continues to be an essential part [2]. Briefly, spintronic materials can be divided into two categories: magnetic metals and magnetic semiconductors. More precisely, magnetic metals can be subdivided into conventional ferromagnetic metals and half-metals (HMs) [3], while magnetic semiconductors can be subdivided into half-semiconductors (HSCs) [4], bipolar magnetic semiconductors (BMSs) [5], and spin-gapless semiconductors (SGSs) [6]. For a spintronic material, a high degree of spin polarization of its free carriers is vitally important on account of that it directly determines the performance of the spintronic devices [7]. Supplementary figure S1 summarizes the schematic density of states (DOSs) of these various spintronic materials. Clearly, conventional ferromagnetic metals can only supply partially spin-polarized carriers due to their low degree of spin polarization at the Fermi level, which greatly limits their practical applications. By contrast, HMs can provide completely spin-polarized carriers because of their specific electronic properties, i.e., one spin channel is metallic while the other is semiconducting. Thus, HMs are viewed as ideal spintronic materials for generating pure spin currents. In addition, magnetic semiconductors, including HSCs, BMSs and SGSs, are also important spintronic materials. HSCs and BMSs have similar electronic properties that both the spin-up and spin-down channels are splitting and have their respective semiconducting or insulating band gaps. The difference is that the valence band maximum (VBM) and conduction band minimum (CBM) of HSCs are fully spin polarized in the same spin direction while those of BMSs possess the opposite spin channels. In SGSs, the spin-split VBM and CBM touch with each other exactly at the Fermi level.

For generating pure spin currents in practical application, these magnetic semiconductors should be further turned into HMs through some external influences, such as external strain [8, 9], electric or magnetic field [4, 7, 10], and carrier doping [5, 9, 11, 12]. In fact, these methods can not only regulate the electronic properties of spintronic materials, but also can tune some two-dimensional (2D) anti-ferromagnetic (AFM) or even nonmagnetic (NM) materials into HMs. For instance, some NM metals and semiconductors, such as $NbS_2$ and $NbSe_2$ nanosheets [13], and edge-doped zigzag hexagonal boron nitride (*h*-BN) nanoribbon [14], can transform into HMs under external strain. Half-metallicity had also been induced in zigzag graphene or *h*-BN nanoribbons by applying an external electric or magnetic field [15, 16]. By using carrier doping, a transition from the AFM to ferromagnetic (FM) state and half-metallicities were achieved in $MnPSe_3$ nanosheet [17], armchair black phosphorene nanoribbon [18], and 2D silicon phosphides [19]. These effective methods provide more options and possibilities for searching new HMs.

On the other hand, with the continuous reduction in the size of microelectronic devices, 2D spintronic materials are eagerly pursued in recent years. Since the discovery of graphene [20], a large number of 2D materials with various structures have emerged [21, 22]. Among them, 2D kagome lattices have been the subject of many studies for decades due to their specific magnetic and electronic properties [23-25]. Very recently, a novel honeycomb-kagome (HK) lattice was proposed in monolayer $Hg_3As_2$ [26] and $Al_2O_3$ [27]. As shown in supplementary figure S2, this HK structure can be viewed as a mixed lattice composed by honeycomb and kagome lattices, and significantly increases the structural diversity of 2D materials. It is known that high chemical diversity and structural complexity will give more opportunities for unique properties. Thus, this novel family of 2D materials with HK lattice quickly attracted many attentions and becomes a new research focus [28-31].

In this work, we proposed a new 2D material with HK lattice, i.e., the $Mg_3C_2$ monolayer, based on the particle-swarm optimization (PSO) algorithm and first-principles calculations. The dynamical and thermal stabilities of this monolayer were confirmed by its phonon spectrum and first-principles molecular-dynamics (FPMD) calculations. The $Mg_3C_2$ monolayer was proved to be an AFM semiconductor at its ground state. Importantly, we demonstrated that a transition from AFM semiconductor to FM half-metal in this material can be induced by either electron or hole doping. The half-metallicity arises from the $2p_z$ orbitals of the C atoms for the electron-doped system, and from the C $2p_x$ and $2p_y$ orbitals for the hole-doping case. In addition, considering that the $Ca_3C_2$, $Sr_3C_2$, and $Ba_3C_2$ monolayers may have the same HK structures as $Mg_3C_2$, we have also commented on their stabilities and electronic properties at the end of this paper.

## 2. Computational Methods

The PSO method within the evolutionary scheme as implemented in the CALYPSO code [32] was employed to predict the low-energy structures of 2D $Mg_3C_2$ monolayers. This method has been used to predict a number of 2D materials successfully [33]. In this PSO simulation, the population size and the number of generation were set to be 50 and 30, respectively. The number of formula units per simulation cell was set to be 1–3, that is, a unit cell containing a total number of atoms of 5, 10 and 15, respectively, was considered.

All first-principles calculations were carried out by using the Vienna Ab initio Simulation Package (VASP) [34]. The ion–electron interaction is represented by the projector augmented plane wave pseudopotential [35] approach and the Perdew–Burke–Ernzerhof (PBE) type generalized gradient approximation [36] is adopted for the exchange–correlation functional. The energy cutoff for plane waves was chosen to 500 eV and the vacuum space along the $z$ direction was set to 15 Å to prevent the interaction between the two neighboring layers. The 2D Brillouin zone integration was sampled with a $23 \times 23 \times 1$ $\Gamma$-centered $k$-point mesh for the unit cell and a $13 \times 13 \times 1$ grid for a larger $2 \times 2$ supercell. Further increasing the number of $k$-points only gave rise to an energy change of less than 0.1 meV. All the atoms in the models were allowed to relax without any constraints and the convergence of force was set to 0.001 eV Å$^{-1}$. To count the electron correlation effects and obtain accurate electronic and magnetic properties of the $Mg_3C_2$ system, more accurate hybrid functional calculations based on the Heyd-Scuseria-Ernzerhof (HSE06) functional [37] were also carried out. The spin−orbital coupling effect was not considered in this study since our test calculations showed that it has little influence on our results.

The phonon spectrum was calculated based on the density functional perturbation theory method as implemented in the phonopy code [38] and the force constant matrix was determined by the VASP. In addition, FPMD simulations, as also performed in the VASP, were employed to study the thermal stability of the $Mg_3C_2$ monolayer. The initial configuration with a $4 \times 4$ supercell was simulated under different temperatures of 300, 500 and 1000 K, respectively. The temperature was controlled by the Nosé-Hoover thermostat [39]. At each temperature, FPMD simulation in *NVT* ensemble lasted for 20 ps with a time step of 2.0 fs.

## 3. Results and discussion

### 3.1. Geometric structure and stability

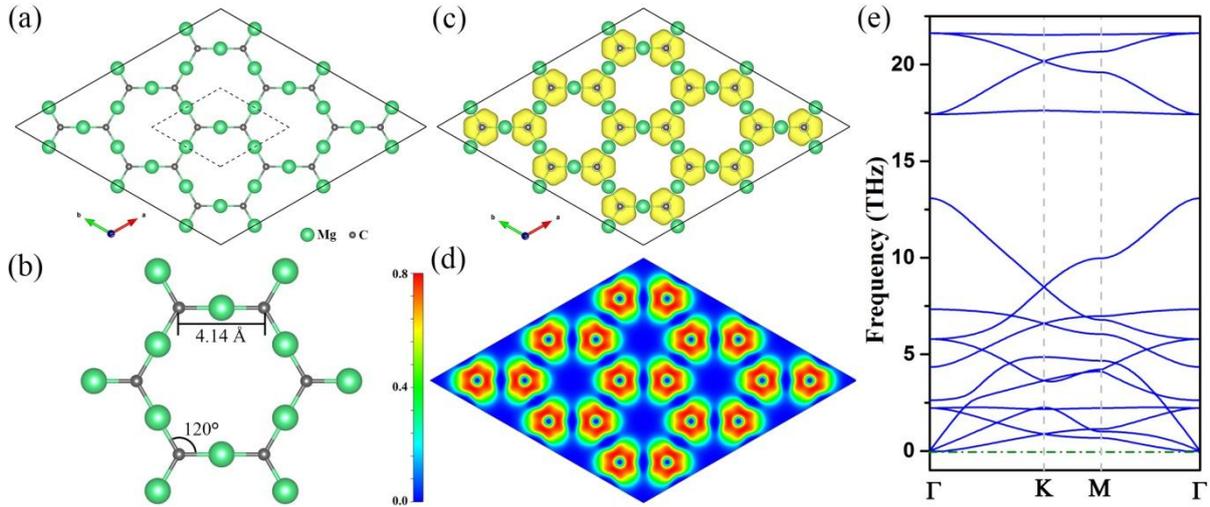

**Figure 1.** (a) Top view of the optimized geometry of $Mg_3C_2$ monolayer with a unit cell labeled by the black dotted line. The green (big) and grey (small) spheres represent Mg and C atoms, respectively. (b) The detailed structure with the marked C-C distance and Mg-C-Mg angle. (c) ELF with an isovalue of 0.5. (d) ELF maps sliced perpendicular to the (001) direction. The color of red and blue refer to the highest (0.80) and lowest value (0) of ELF, respectively. (e) Phonon spectrum.

The most energetically stable structure of $Mg_3C_2$ monolayer predicted by the PSO simulation is plotted in figures 1(a) and (b). Clearly, this $Mg_3C_2$ monolayer has a planar HK lattice configuration with the *P6/mmm* symmetry. As shown in figure 1(a), its primitive cell contains three Mg and two C atoms with the optimized lattice constants being $a = b = 7.16$ Å. In this planar monolayer, three neighboring Mg atoms are located around each C atom with the bond angle C–Mg–C of $120°$ and Mg–C bond length of 2.07 Å (figure 1(b)). Besides, other structures of $Mg_3C_2$ monolayer predicted by the PSO structure search, for instance, the *pmmm*-Mg$_3$C$_2$ shown in supplementary figure S3, have higher energy and evident imaginary frequencies. Therefore, in the next sections, we only study the Mg$_3$C$_2$ monolayer with HK lattice which has the global minimum structure.

To reveal the bonding nature of this monolayer, we calculated the electron localization function (ELF) to analyze its electron distributions. As a helpful method for classifying chemical bonds, the ELF can be visually described in the form of isosurface in real space with isovalues between 0 and 1. The region with the smaller isovalue implies the area with the lower electron density, and the region with 0.5 represents the area with a homogeneous electron gas. As shown in figure 1(c), electrons are mainly distributed and localized at the region around C atoms, showing the evident characteristic of ionic bond. In addition, the middle of the Mg–C bond still possesses some electrons, indicating that the Mg–C bonds also have the feature of a weak covalent bond. This result can be further evidenced by the ELF map sliced perpendicular to the (001) direction of the Mg$_3$C$_2$ monolayer (Figure 1(d)). To quantify the bonding description, Bader charges of Mg and C atoms were calculated. According to the Bader charge analysis [40, 41], Mg and C atoms in this monolayer respectively possess +1.43 and -2.14 |*e*| charge. Namely, the three Mg atoms lost 4.28 ELECTRONS to the two C atoms in one unit cell due to their different electronegativity. It is expected that the strong ionic and weak covalent electron states can stabilize the 2D framework of Mg$_3$C$_2$.

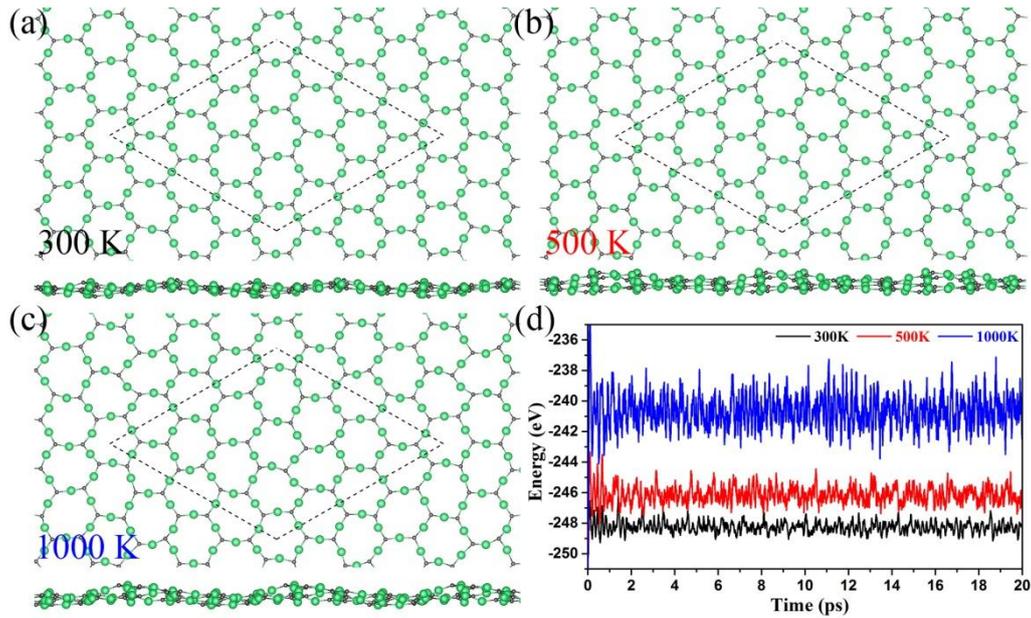

**Figure 2.** Thermal stability of the $Mg_3C_2$ monolayer. Snapshots for the equilibrium structures at the end of 20 ps FPMD simulations under the temperatures of (a) 300, (b) 500 and (c) 1000 K. The black dashed lines denote the $4 \times 4$ supercell used in the FPMD simulations. (d) The fluctuations of total energies of $Mg_3C_2$ system with respect to FPMD steps at 300, 500, and 1000 K.

To confirm this expectation and evaluate the stability of the $Mg_3C_2$ monolayer, we first calculated its phonon dispersion curves. As shown in figure 1(e), no imaginary mode is observed in the phonon spectrum, confirming the dynamical stability of the $Mg_3C_2$ monolayer. We then performed FPMD simulations using a $4 \times 4$ supercell to further evaluate its thermal stability. The initial configuration was simulated at different temperatures of 300, 500 and 1000 K with a time step of 2 fs. As shown in figures 2(a)–(c), the snapshots of the geometry structure at the end of 20 ps simulations clearly reveal that this monolayer can maintain its structural integrity except for some thermal fluctuations. Unsurprisingly, the thermal fluctuation increases with the rising temperature (Figure 2(d)). Nonetheless, the total energy of the simulated system only fluctuates around a certain constant magnitude under the certain temperature. These results demonstrate the remarkable thermal stability of the Mg$_3$C$_2$ monolayer.

### 3.2. Electronic structures and magnetic properties

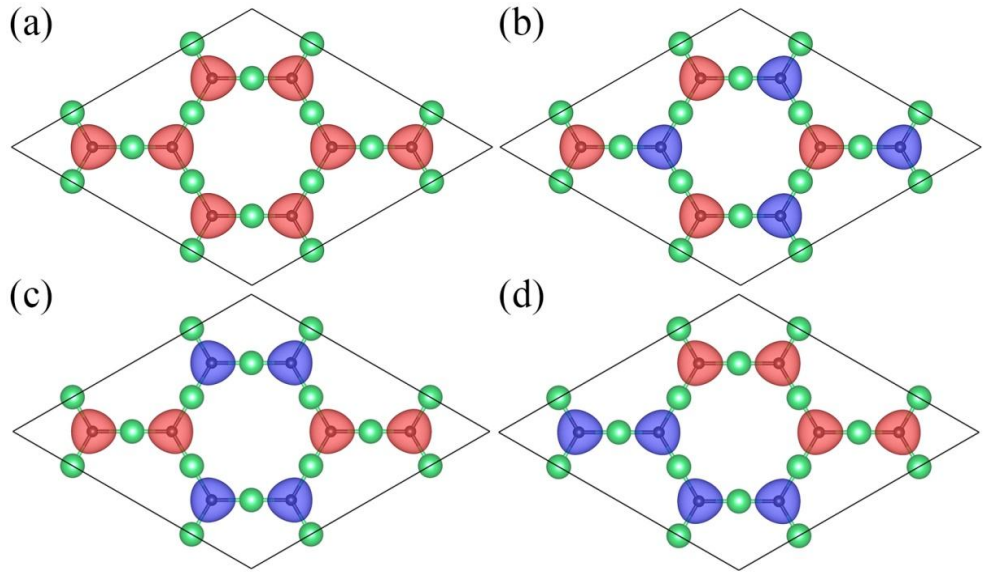

**Figure 3.** The spatial distribution of the spin-polarized electron density of the Mg$_3$C$_2$ monolayer with (a) FM, (b) AFM1, (c) AFM2 and (d) AFM3 states. The isovalue is set to 0.02 e Å$^{-3}$. Red and blue colors indicate the positive (spin-up state) and negative (spin-down state) values, respectively.

To study the electronic property of the Mg$_3$C$_2$ monolayer, its preferred magnetic ground state should be first determined. Hence, we calculated the total energies of the Mg$_3$C$_2$ monolayer with FM, AFM, and NM states. The configurations of FM state and three possible AFM (AFM1, AFM2 and AFM3) states are shown in figure 3. The calculated energies reveal that the AFM1 state is the most energetically stable magnetic state with the energy of 19.6, 40.3, 62.9 and 526.5 meV per unit cell lower than that of AFM2, AFM3, FM and NM states, respectively. At this ground state, the magnetic moments are mainly localized around C atoms and all of the nearest C neighbors have antiparallel spins (figure 3(b)). These results indicate that the Mg$_3$C$_2$ monolayer is AFM coupling under its natural condition. Notice that all the AFM states we discussed in the following sections are referring to the AFM1 state due to its lower energy.

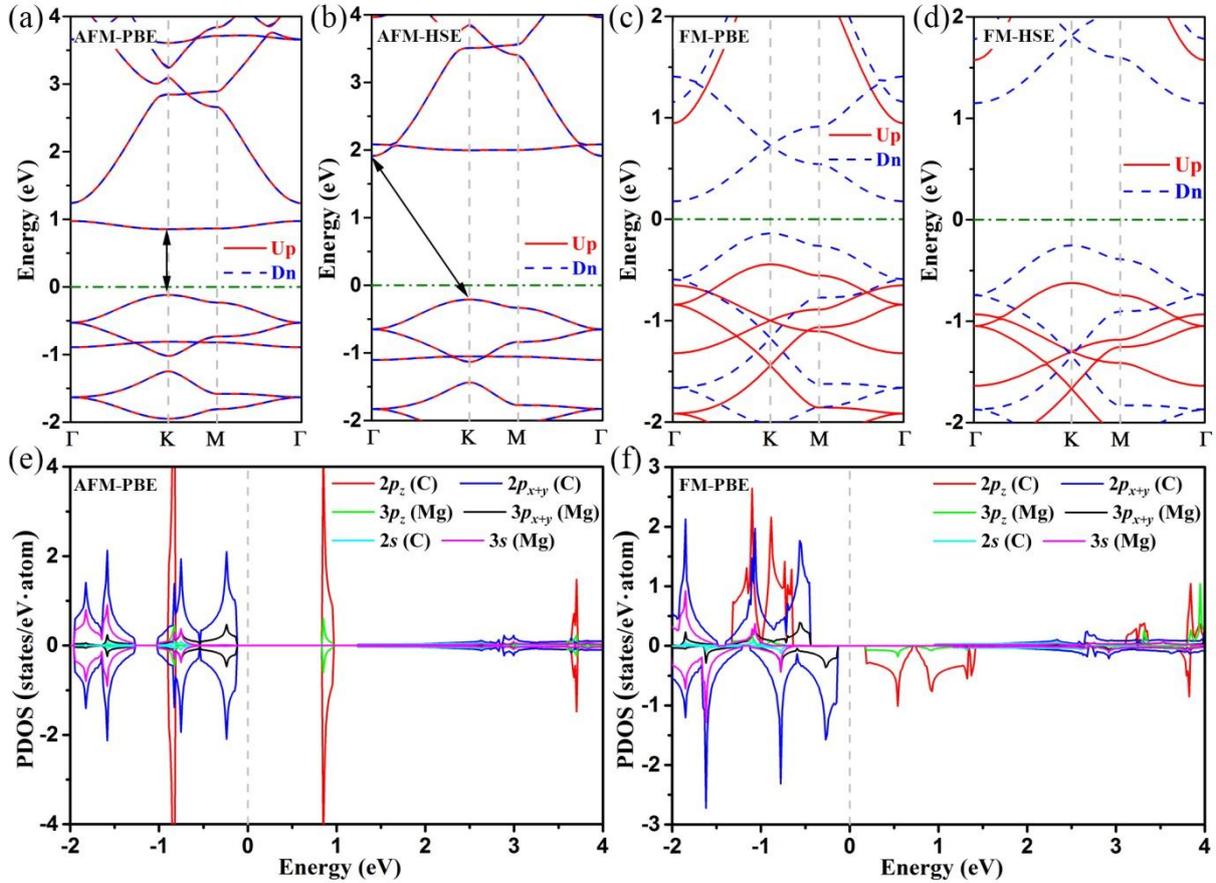

**Figure 4.** Spin-resolved band structures within (a) PBE functional and (b) HSE06 functional for the AFM state of the $Mg_3C_2$ monolayer. (c) and (d) are respectively similar to (a) and (b), but for the FM state. Parts (e) and (f) are PDOSs of the $Mg_3C_2$ monolayer with AFM and FM states, respectively. The Fermi energy is set to zero.

To further study the electronic property of the $Mg_3C_2$ monolayer, its spin-resolved band structures were respectively calculated based on PBE and HSE06 functionals (figures 4(a) and (b)). Clearly, energy bands for the spin-up and spin-down electrons are completely degenerate and show semiconducting behaviors. The band structure calculated within the PBE functional demonstrates that the $Mg_3C_2$ monolayer displays a direct band gap (0.97 eV) at the $K$ point. However, within the HSE06 functional, the bottom of conduction band shifts up, leading to an indirect band gap (2.12 eV) from the $\Gamma$ to the $K$ point. Furthermore, as displayed in figures 4(a) and (b), the valence bands have little dispersion with narrow bandwidths, indicating the localized character of electrons in this system. Notably, these flat bands result in sharp van Hove singularities with large DOS, as shown in figure 4(e). It was reported that the sharp van Hove singularities in some 2D materials can lead to a magnetic instability and induce a magnetic transition to FM state by some external influence [18, 42, 43].

Thus, we also calculated the band structures of the $Mg_3C_2$ monolayer with FM state using the PBE and HSE06 functionals, respectively. As shown in figures 4(c) and (d), both the band structures present the typical characteristic of HSCs [4], i.e., the VBM and CBM are fully spin polarized in the same direction (spin-down direction). The spin-down and spin-up channels respectively have indirect band gaps of 0.32 and 1.39 eV from the $\Gamma$ to the $K$ point within the PBE functional (figure 4(c)), while the values are 1.40 and 2.21 eV for the HSE06 functional (figure 4(d)). Moreover, projected DOS (PDOS) (figures 4(e) and (f)) analysis shows that the VBM mainly originates from the $2p_x$ and $2p_y$ orbitals of the C atoms, while the CBM is mainly contributed by the C $2p_z$ orbitals for both AFM and FM states.

### 3.3. Manipulation of magnetic state

Attracted by the exotic electronic property of the $Mg_3C_2$ monolayer in its FM state, we naturally studied how to induce a magnetic transition from its AFM to FM state. External strain, electric field, and carrier doping are three convenient methods for regulating the electronic properties of 2D materials. Thus, we turn to explore these external influences on the electronic properties of the

Mg$_3$C$_2$ monolayer in the following sections. To assess the possibility of magnetic transition, we computed the total energy difference per unit cell between the FM and AFM states ($E_{FM}$ - $E_{AFM}$) under different cases. Positive energy difference indicates the intrinsic AFM state of this system surviving, while negative ones demonstrate the magnetic ground state changing to FM state.

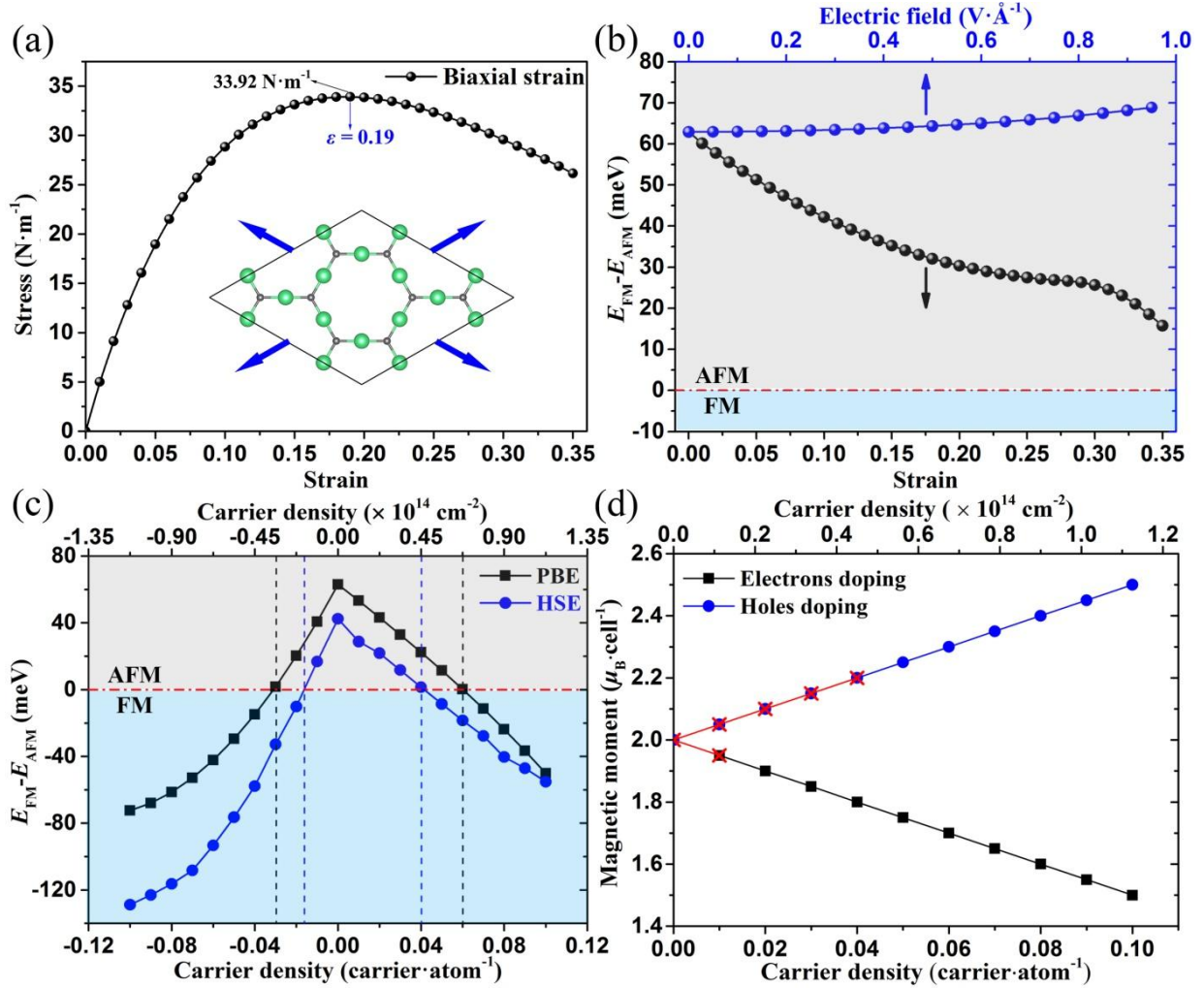

**Figure 5.** (a) Calculated stress-strain relationships of the Mg$_3$C$_2$ monolayer under biaxial tensile strains. The inset shows the primitive cell of Mg$_3$C$_2$ with the two strain orientations. (b) The $E_{FM}$ - $E_{AFM}$ with respect to tensile strains and external electric fields. (c) The $E_{FM}$ - $E_{AFM}$ with respect to carrier doping calculated with PBE and HSE06 functional. The negative and positive values on the horizontal ordinate are for electron and hole doping, respectively. The zero energy difference is marked by the red dot-dashed lines. (d) The doping dependence of the magnetic moment of a Mg$_3$C$_2$

unit cell. The points with and without red crosses indicate the AFM and FM ground states calculated within the HSE06 functional, respectively.

First, we applied homogeneous biaxial strains to the $Mg_3C_2$ monolayer by changing its lattice parameters. Since 2D materials are prone to wrinkle under lateral compression, we consider only the tensile strains here. This tensile strain is defined as $\varepsilon = (a - a_0)/a_0$, where $a$ and $a_0$ are the lattice parameters with and without deformation, respectively. As shown in figure 5(a), the stress-strain curve we calculated clearly reveals that this monolayer can sustain a biaxial strain up to 0.19 and shows a superior flexibility. Thus, a wide range of strains (0–0.35) we applied is large enough for this material. In addition, we also studied the effects of external electric fields with the strength ranging from 0.0 to 0.95 V $Å^{-1}$. These electric fields were applied along the $z$-axis which is perpendicular to the $Mg_3C_2$ monolayer. Figure 5(b) presents the $E_{FM}$ - $E_{AFM}$ values under different biaxial tensile strains and electric fields, respectively. Clearly, all $E_{FM}$ - $E_{AFM}$ values are positive, indicating that the AFM state is always the magnetic ground state of the $Mg_3C_2$ monolayer under biaxial tensile strains or external electric fields in a rational range.

Then, we turn to examine the carrier doping effects on the electronic and magnetic properties of this system by introducing additional electrons or holes. The range of doping concentrations applied in this work is from 0 to 0.1 carriers (electrons or holes) per atom. The maximal doping concentration of 0.1 carriers per atom, corresponding to the carrier density of $1.12 \times 10^{14}$ $cm^{-2}$, can be easily realized in experiment. In fact, a carrier density modulation has already been experimentally achieved to the order of $10^{15}$ $cm^{-2}$ for 2D materials by employing an electrolytic gate [44-46]. Remarkably, although the $Mg_3C_2$ monolayer can retain its AFM ground state under a small amount of carrier doping, the values of $E_{FM}$ - $E_{AFM}$ decrease rapidly with the increase of either electron or hole doping (figure 5(c)). When the carrier doping concentration reaches about 0.03 electrons per atom or 0.06 holes per atom (corresponding to 3.38 $\times 10^{13}$ cm$^{-2}$ and 6.76 $\times 10^{13}$ cm$^{-2}$), respectively, the value of $E_{FM}$ - $E_{AFM}$ decreases to zero. And this value turns negative with the doping concentration further increasing, demonstrating that the FM rather than AFM coupling is energetically favorable in these carrier-doping cases. That is, the 2D $Mg_3C_2$ system undergoes a magnetic transition from AFM to FM state with increasing the carrier doping above a moderate concentration.

Considering that the HSE06 functional, which includes the accurate Fock exchange, usually performs much better in counting the electron correlation effects than the DFT and DFT+U methods [47-49], we thus employ it to recheck the above magnetic transition of the $Mg_3C_2$ system. As shown in figure 5(c), the variation trend of the $E_{FM}$ - $E_{AFM}$ with respect to carrier doping calculated within HSE06 functional is as the same as that calculated within PBE functional. Furthermore, the magnetic transition occurs at a lower carrier concentration, i.e., about 0.016 electrons per atom and 0.04 holes per atom (corresponding to 1.80 $\times 10^{13}$ cm$^{-2}$ and 4.50 $\times 10^{13}$ cm$^{-2}$, respectively). These results verify the magnetic transition from AFM to FM state induced by carrier (electron or hole) doping in this 2D $Mg_3C_2$ system. Moreover, as shown in figure 5(d), one can note that the magnetic moment of the carrier doped system decreases linearly with increasing electron doping, while increases with the increase of hole doping. The ratio of magnetic moment to the doped electrons or holes is 1 $\mu_B$/electron or 1 $\mu_B$/hole. Namely, the doped electrons and holes are completely polarized, implying that the doped $Mg_3C_2$ system may present a half-metallic behavior.

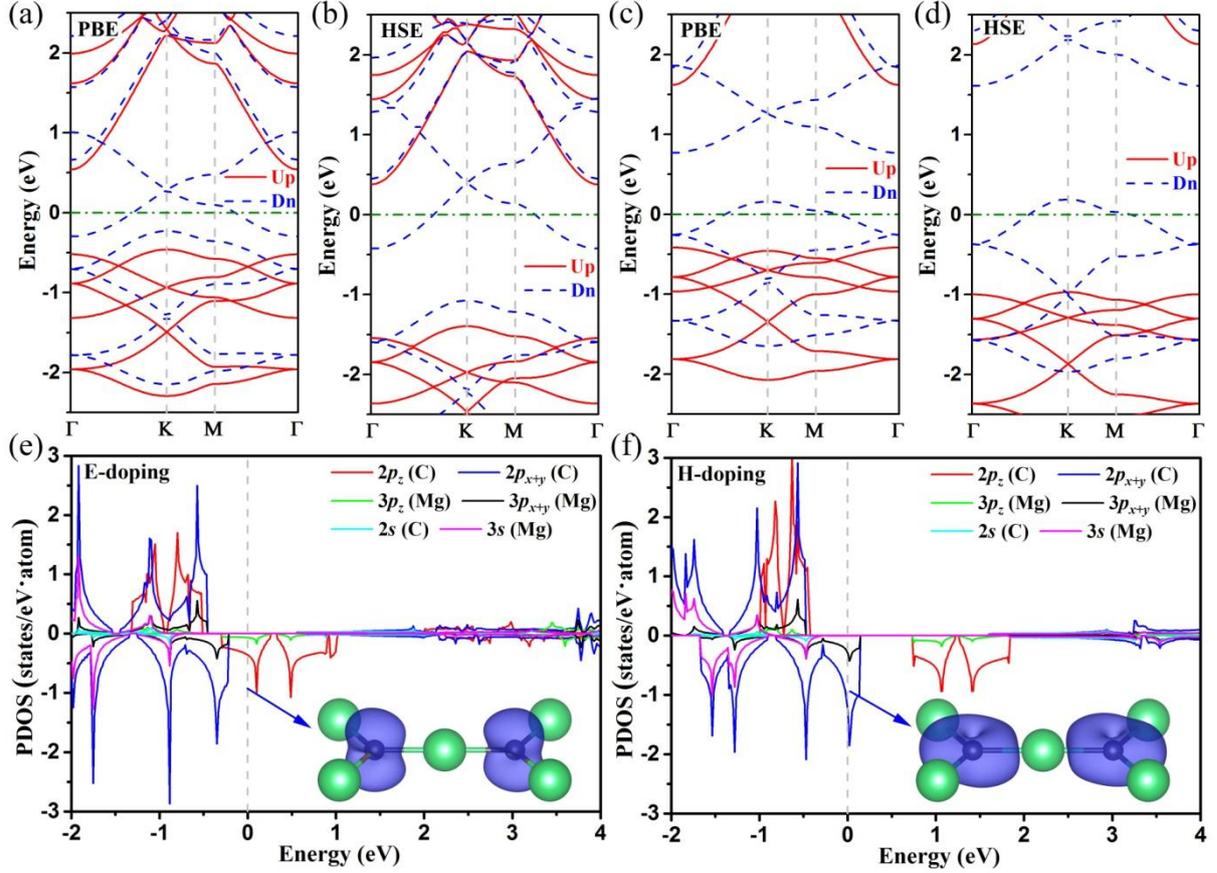

**Figure 6.** (a) and (b) are band structures of the Mg$_3$C$_2$ monolayer with the doping concentration of 0.1 electrons per atom calculated within PBE and HSE06 functional, respectively. (c) and (d) are similar to (a) and (b), respectively, but for the doping concentration of 0.1 holes per atom. Parts (e) and (f) are spin-resolved PDOSs of the Mg$_3$C$_2$ monolayer with the doping concentration of 0.1 electrons per atom and 0.1 holes per atom, respectively. The Fermi energy is set to zero and the insets are charge distribution of the energy band across the Fermi level plotted with an isovalue of 0.01 e Å$^{-3}$.

We next clarify the electronic structure of the doped Mg$_3$C$_2$ system after magnetic phase transformation. Figure 6 presents the spin-resolved band structures and PDOSs of the Mg$_3$C$_2$ system with the doping concentration of 0.1 carriers per atom. In comparison with the band structure for the

FM state of this monolayer without carrier doping (figures 4(c) and (d)), one can see clearly that the Fermi levels either shifts up to the conduction bands due to the influences of electron doping (figures 6(a) and (b)) or shifts down to the valence bands in the case of hole doping (figures 6(c) and (d)). Notably, the conduction and valence bands near the Fermi level are contributed by carriers with spin-down states, while the spin-up channel presents a semiconducting behavior with a gap of 1.06 eV for electron doping case and 2.03 eV for hole doping case. These values are calculated to be 1.92 and 3.13 eV, respectively, within the HSE06 functional. These features of the band structure indicate that the doped $Mg_3C_2$ system is indeed a HM, and thus can provide completely spin-polarized currents. Furthermore, the states around the Fermi level are mainly contributed by the $2p_z$ orbitals of the C atoms for the electron-doped $Mg_3C_2$ monolayer (figure 6(e)), but by the C $2p_x$ and $2p_y$ orbitals for the hole-doped case (figure 6(f)). This result can be further verified and visualized by the spatial charge distribution profiles of the energy bands across the Fermi level [insets of figures 6(e) and (f)]. It is evident that the shape of the spatial charge on the C atoms has distinctly characteristic of $2p_z$ (out-of-plane) and $2p_x$ ($2p_y$) (in-plane) orbitals for the electron-doped and hole-doped $Mg_3C_2$, respectively.

Finally, we have also examined the stabilities and electronic properties of the $Ca_3C_2$, $Sr_3C_2$, and $Ba_3C_2$ monolayers with the same HK lattice as $Mg_3C_2$. These 2D structures can be viewed as replacing the Mg atoms in the $Mg_3C_2$ monolayer with larger atoms in the same element group, i.e., calcium, strontium, or barium. As shown in supplementary figure S4, the $Ca_3C_2$ monolayer presents a half-metallic behavior at its ground state. However, its rather soft phonon spectrum with small imaginary frequency in the transverse acoustical phonon branch reveals that this 2D material can hardly exist in a free-standing state. The phonon spectra of the $Sr_3C_2$ and $Ba_3C_2$ monolayers show evident imaginary frequencies, which can be seen from supplementary figure S5, denoting their instabilities.

## 4. Conclusions

In conclusion, we have proposed a new 2D material with HK lattice, i.e., the $Mg_3C_2$ monolayer, and have confirmed its dynamical and thermal stabilities by phonon dispersion curves and FPMD simulations, respectively. Based on first-principle calculations, we found that this monolayer is an AFM semiconductor at its ground state. More importantly, we further demonstrated a magnetic transition from AFM to FM state in this monolayer induced by electron or hole doping. In particular, the doped $Mg_3C_2$ monolayer is a HM with completely spin-polarized carriers. In addition, the half-metallicity arises from the $2p_z$ orbitals of the C atoms for the electron-doped system, and from the C $2p_x$ and $2p_y$ orbitals for the hole-doping case. Our findings demonstrate the potential utilization of the $Mg_3C_2$ monolayer in 2D spintronic devices with electrically controllable spin currents. We hope that our study will stimulate further experimental effort in synthesizing this 2D material with HK lattice.

## Acknowledgments

This work was supported by the State Key Program for Basic Research of China (Grant Nos. 2014CB921102 and 2017YFA0206304), the National Natural Science Foundation of China (Grant Nos. 51572122, 11304096 and 11674144), and the Natural Science Foundation of Shandong Province (Grant Nos. JQ201602 and ZR2017JL007). We are grateful to the High Performance Computing Center of Nanjing University for doing the numerical calculations in this paper on its blade cluster system.